\documentclass{article}
\usepackage{emulateapj}
\usepackage{apjfonts}

\begin{document}

\newcommand{\figschem}{1}
\newcommand{\figdark}{2}
\newcommand{\figgeo}{3}
\newcommand{\figlimb}{4}
\newcommand{\tabobs}{1}

\submitted{To appear in The Astronomical Journal}

\title{Measurements of the Diffuse Ultraviolet Background and the Terrestrial
Airglow with the Space Telescope Imaging Spectrograph$^{1}$}


\author{Thomas M. Brown$^{2,4}$, Randy A. Kimble$^{2,6}$, 
Henry C. Ferguson$^{3}$,
Jonathan P. Gardner$^{2,7}$, Nicholas R. Collins$^{2,5,8}$,
Robert S. Hill$^{2,5,9}$}

\noindent 
\begin{abstract}

Far-UV observations in and near the Hubble Deep Fields demonstrate
that the Space Telescope Imaging Spectrograph (STIS) can potentially
obtain unique and precise measurements of the diffuse far-ultraviolet
background.  Although STIS is not the ideal instrument for such
measurements, high-resolution images allow Galactic and extragalactic
objects to be masked to very faint magnitudes, thus ensuring a
measurement of the truly diffuse UV signal. The programs we have
analyzed were not designed for this scientific purpose, but would be
sufficient to obtain a very sensitive measurement if it were not for a
weak but larger-than-expected signal from airglow in the
STIS 1450--1900~\AA\ bandpass.  Our analysis shows that STIS far-UV
crystal quartz observations taken near the limb during orbital day can
detect a faint airglow signal, most likely from \ion{N}{1} $\lambda
1493$~\AA, that is comparable to the dark rate and inseparable from
the far-UV background.  Discarding all but the night data from these
datasets gives a diffuse far-ultraviolet background measurement of
$501 \pm 103$ ph cm$^{-2}$ sec$^{-1}$ ster$^{-1}$ \AA$^{-1}$, along a
line of sight with very low Galactic neutral hydrogen column ($N_{HI}
= 1.5\times10^{20}$ cm$^{-2}$) and extinction ($E(B-V)=0.01$ mag).  This
result is in good agreement with earlier measurements of the far-UV
background, and should not include any significant contribution from
airglow.  We present our findings as a warning to other
groups who may use the STIS far-UV camera to observe faint extended
targets, and to demonstrate how this measurement may be properly
obtained with STIS.

\end{abstract}


\section{INTRODUCTION} \label{secintro}

Over the past several decades, the far-ultraviolet background (FUVBG)
has been the subject of many studies and the center of considerable
controversy (for a review and an example of such debates, see Bowyer
1991\markcite{B91}; Henry 1991\markcite{H91}).  The FUVBG is defined
as the diffuse astronomical flux incident upon the Earth at
wavelengths of 912--2000~\AA, thus excluding sources such as
terrestrial airglow and Galactic stars.  Because the zodiacal light
decreases dramatically below 2500~\AA, and because the diffuse UV
background is so faint, O'Connell (1987\markcite{O87}) has noted that
this spectral region offers a window of very low sky background for
observations of objects with faint surface brightness.  We have been
utilizing this spectral window for UV observations of faint extended
sources, but these data also provide insight into the FUVBG itself.

A number of FUVBG measurements have shown a correlation between
the FUVBG intensity and the Galactic neutral hydrogen column density
($N_{HI}$), leading to the conclusion that a significant fraction
of the FUVBG must be Galactic in origin (see, e.g., the relatively
recent data set of Hurwitz, Bowyer, \& Martin 1991\markcite{HBM91}).
In a 1400--1850~\AA\ bandpass, extrapolating to
\medskip

{\small
$^1$Based on observations with the NASA/ESA Hubble Space
Telescope obtained at the Space Telescope Science Institute, which is
operated by AURA, Inc., under NASA contract NAS~5-26555.

$^2$Laboratory for Astronomy \& Solar Physics, Code 681, Goddard Space
Flight Center, Greenbelt, MD 20771

$^3$STScI, 3700 San Martin Drive,
Baltimore, MD 21218.  ferguson@stsci.edu

$^4$NOAO Research Associate. tbrown@pulsar.gsfc.nasa.gov

$^5$Raytheon ITSS Corp. 

$^6$kimble@ccd.gsfc.nasa.gov

$^7$gardner@harmony.gsfc.nasa.gov 

$^8$collins@zolo.gsfc.nasa.gov

$^9$bhill@virgil.gsfc.nasa.gov
} 

\noindent
 $N_{HI}=0$ gives an FUVBG intensity of $\sim$300 ph cm$^{-2}$
sec$^{-1}$ ster$^{-1}$ \AA$^{-1}$; this intensity increases to
$\sim$1200 ph cm$^{-2}$ sec$^{-1}$ ster$^{-1}$ \AA$^{-1}$ with
increasing $N_{HI}$, but saturates before high column densities are
reached (Hurwitz et al.\ 1991\markcite{HBM91}; Bowyer
1991\markcite{B91}).  Curiously, measurements below Ly-$\alpha$
may indicate an upper limit of 30--100 ph cm$^{-2}$ sec$^{-1}$ ster$^{-1}$
\AA$^{-1}$ (Henry 1991\markcite{H91}; Murthy et al.\
1999\markcite{M99}), although this measurement has been contested
(Edelstein, Bowyer, \& Lampton 2000\markcite{E00}).  
Herein lies the controversy: a step in the FUVBG
at Ly-$\alpha$ could imply an extragalactic origin for the $N_{HI}=0$
component, such as redshifted Ly-$\alpha$ emission from an
intergalactic medium (IGM) (Henry 1991\markcite{H91}); demonstration
of such an origin of the FUVBG would have extremely important
consequences, as canonical values for IGM parameters predict a
negligible contribution to the FUVBG (e.g., Jakobsen
1993\markcite{J93}). On the other hand, others maintain that the
dominant contribution to the $N_{HI}=0$ component is mainly
dust-scattered Galactic light (Bowyer 1991\markcite{B91}); under that
assumption, the residual dust at $N_{HI}=0$ that is needed to produce
the FUVBG corresponds to $E(B-V)=0.015$~mag (Hurwitz et al.\
1991\markcite{HBM91}).  We note that we cannot directly discriminate
between an extragalactic and Galactic origin of the FUVBG, as we are
measuring the FUVBG longward of Ly-$\alpha$, but our measurements have
all been taken along lines of sight with extremely low $N_{HI}$ column
{\it and} low dust extinction (see Table \tabobs).  New maps of
Galactic extinction (Schlegel, Finkbeiner, \& Davis
1998\markcite{SFD98}) are based on a direct measurement of the IR
emission from the dust, instead
of inferring the extinction from dust-to-hydrogen ratios; these new
dust maps can potentially yield new insight into the FUVBG debate.

The Hubble Space Telescope (HST) currently employs one instrument with
significant sensitivity in the far-UV: the Space Telescope Imaging
Spectrograph (STIS) (Woodgate et al.\ 1998\markcite{W98}).  However,
STIS is not an ideal instrument for measuring the FUVBG: the
throughput in the crystal quartz bandpass 

\newpage

\parbox{5.5in}{
{\sc Table \tabobs:} Far-UV Data

\begin{tabular}{clccccrcc}
\tableline
Field    & Exposure & RA             & Dec                            & $l$    & $b$     & coverage        & $N_{HI}$$^a$ & $E(B-V)^b$ \\
ID       & (sec)    & (J2000)        & (J2000)                        & (deg)  & (deg)   & (square arcsec) &  cm$^{-2}$   & (mag) \\
\tableline
HDFN-Fol & 124330   & $12^h36^m44^s$ & $\rm +62^o12\arcmin 12\arcsec$ & 125.908 & +54.840 & 3659 & 1.51$\times 10^{20}$ & 0.01 \\
HDFN-Par &  17315   & $12^h35^m41^s$ & $\rm +62^o10\arcmin 38\arcsec$ & 126.124 & +54.853 & 1003 & 1.51$\times 10^{20}$ & 0.01 \\
HDFS-Pri &  52124   & $22^h33^m38^s$ & $\rm -60^o33\arcmin 29\arcsec$ & 328.158 & -49.274 &  878 & 2.39$\times 10^{20}$ & 0.03 \\
\tableline
\end{tabular}

$^a$ Dickey \& Lockman (1990)\markcite{DL90}.\\
$^b$ Schlegel et al.\ (1998)\markcite{SFD98}.
}

\bigskip

\noindent
is not very high (peaking at
573 cm$^2$ effective area), the pixels and field of view are small
(respectively $0.025 \arcsec$ and $25\arcsec$), and the dark
background is spatially and temporally variable (but very low).
Nonetheless, we have found that the deepest imaging programs can yield
interesting measurements of the diffuse UV background, because deep,
high-resolution, coincident imaging at longer wavelengths can provide
a mask for all objects brighter than $\sim$ 29 mag in the optical.
Because we can mask very faint galaxies, we can determine how much of
the FUVBG is truly diffuse.  In this paper, we demonstrate the
measurements of the FUVBG that could be made using the STIS far-UV
camera.  The programs we analyze here, from observations in and near
the Hubble Deep Fields (HDFs), were not intended for measurements of
the FUVBG, but they indicate that an HST program could be tailored to
characterize the FUVBG with excellent statistics.

\section{OBSERVATIONS} \label{secobs}

There are three programs with deep coincident far-UV and optical
imaging that can potentially yield interesting  measurements of the
FUVBG: STIS followup imaging (General Observer Program 7410) of the
HDF North (HDFN) (Williams et al.\ 1996\markcite{W96}),
STIS parallel imaging (Guaranteed Time Observer 
Programs 7920 \& 7921) during Near Infrared Camera and Multi Object
Spectrometer (NICMOS) observations of the HDFN (Thompson et al.\
1999\markcite{T99}), and primary STIS imaging (Gardner et al.\
2000a\markcite{G00a}) from the HDF South (HDFS) campaign (Williams et
al.\ 2000\markcite{W00}).  We will hereafter refer to each of these
fields as HDFN-Fol, HDFN-Par, and HDFS-Pri.  Each of these fields has
optical imaging to a 3$\sigma$ limiting AB mag of $\gtrsim 29$.  The
far-UV data for these fields are summarized in Table~\tabobs.  Note
the $N_{HI}$ and extinction along each line of sight are quite low;
because the FUVBG intensity correlates with $N_{HI}$, we would
expect measurements of the FUVBG to be near the minimum of
$\sim$300 ph cm$^{-2}$ sec$^{-1}$ ster$^{-1}$ \AA$^{-1}$
(Bowyer 1991\markcite{B91}; Hurwitz et al.\ 1991\markcite{HBM91}).

The UV imaging in each of these programs utilized the STIS far-UV
camera with the crystal quartz filter (F25QTZ).  This camera employs a
multianode microchannel array (MAMA) instead of a charge-coupled
device (CCD); there is no read noise, and very little sensitivity to
cosmic rays.  The minimum dark rate is 6.6$\times 10^{-6}$ cts
sec$^{-1}$ pix$^{-1}$. The bandpass spans 1450--1900~\AA, negating the
signal from terrestrial Ly-$\alpha$ and \ion{O}{1} $\lambda
1301$~\AA.  Because the long wavelength cutoff of this bandpass is due
to the rapidly dropping sensitivity of the detector's CsI
photocathode, red leak is also negligible.  A full description of the
instrument can be found in Woodgate et al.\ (1998\markcite{W98}) and
Kimble et al.\ (1998\markcite{K98}).
The STIS photometric calibration is reliable at the 0.15 mag level,
according to the STScI documentation (Baum et al.\ 1998\markcite{B98}).

Before our analysis, the lack of other known strong airglow
lines in this bandpass at the HST altitude (600 km) implied
that the airglow signal should be at least an order of magnitude 
\\ \\

\vskip 1.15in

\noindent
below the
minimum dark rate.  Understandably, the spectral sky templates used
for the STScI Exposure Time Calculator also make the same assumption.
However, we have found that the sky signal can reach levels that are a
significant fraction of the dark signal, if observations are made with HST 
near the limb during orbital day, as we discuss in \S \ref{secanal}.

Furthermore, the mean temperature of the far-UV camera has increased
since STIS commissioning, and the dark background increases as a
function of temperature.  This increase appears as a ``glow'' that is
strongest in the upper left-hand quadrant of the detector
(Figure~\figschem), where the dark rate can be 20 times higher than
the minimum rate.  Because the dark rate in this part of the detector
is highly variable, we excluded this region during our analysis, thus
rejecting over 75\% of the available detector area.  A cooler for the
STIS camera should become available in Cycle 10, which would negate
this problem and allow much more detector area for this measurement.
A small portion of the detector is occulted by the aperture mask,
appearing as a strip along the bottom $\sim$15 rows in STIS images;
this region can be used to determine the dark rate in the exposed
portions of the detector, away from the glow, as we discuss in \S
\ref{secdark}. \\

\bigskip

\hskip -0.2in
\parbox{3.0in}{\epsfxsize=3.5in \epsfbox{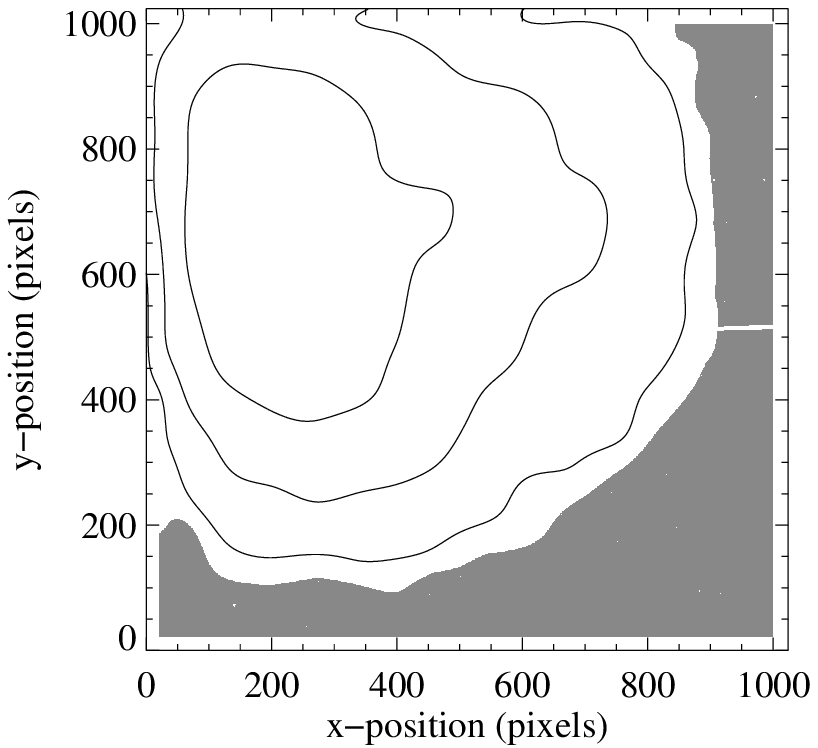}} \vskip 0.1in

\centerline{\parbox{3.5in}{\small {\sc Fig.~\figschem--} A schematic
of the STIS images used for measuring the FUVBG.  When there is a
strong dark background, its shape follows the contours shown (contour
levels are at $2\times$, $5\times$, and $10\times$ the lowest dark
rate of 6.6$\times 10^{-6}$ cts sec$^{-1}$ pix$^{-1}$).  The grey
shading shows portions of the detector free from dark glow, hot
pixels, edge effects, and repeller wire occultation; this portion of the
detector is thus suitable for measuring the FUVBG.  }}
\addtocounter{figure}{1}

\bigskip

\subsection{HDFN Followup Observations (HDFN-Fol)} \label{sechdf}

The original HDFN program produced observations of an undistinguished
field at high Galactic latitude with the Wide Field Planetary Camera 2
(WFPC2), employing the F300W, F450W, F606W, and F814W filters (see
Williams et al.\ 1996\markcite{W96} for a complete description).  A
Cycle 7 General Observer Program (No. 7410) obtained observations of a
portion of the field with the STIS far-UV and near-UV cameras.  A
total of 64 exposures spanning 124330 sec were obtained during a two
year period (1997 December to 2000 February).  The six STIS positions
cover a roughly rectangular region of sky centered at
RA(J2000)=$12^h36^m44^s$ and Dec(J2000)=$\rm 62^o12\arcmin 12\arcsec$,
with an area of 3659 square arcsec, and falling completely within the
WF4 CCD HDFN image (which has an exposed area of 5926 square arcsec).
Small dithers ($\sim$ 10 MAMA pixels) were employed between frames at
each position, to smooth out small-scale changes in detector
characteristics.  Some of these STIS exposures were taken completely
during orbital night, and all were taken near the limb (the
zenith angle in these exposures ranged from 82--95$\rm ^o$).  As we
discuss in \S \ref{secanal}, the day exposures are contaminated by a
faint sky signal that is inseparable from the FUVBG.  The presumably
uncontaminated night exposures were all taken in the two most northern
positions, centered at RA(J2000)=$12^h36^m43^s$ and Dec(J2000)=$\rm
62^o12\arcmin 34\arcsec$.

\subsection{NICMOS HDFN Parallel Observations (HDFN-Par)} \label{secpar}

In 1998 January, Thompson et al.\ (1999\markcite{T99}) observed the
original HDFN field with NICMOS, and STIS performed parallel imaging on
a field $\sim 8.5\arcmin$ away.  The exposure time for each frame was
short, as a result of the parallel
observing procedure.  STIS observed in four imaging modes: far-UV
(F25QTZ), near-UV (F25QTZ), clear CCD (50CCD), and long-pass CCD
(F28X50LP), and one low-resolution near-UV spectroscopic mode
(G230LB).  The clear CCD and far-UV exposures each covered a roughly
square region of sky, with respective areas of 3546 square arcsec and
1003 square arcsec, both centered at RA(J2000)=$12^h35^m41^s$ and
Dec(J2000)=$\rm 62^o10\arcmin 38\arcsec$.  Dithers between far-UV
frames ranged up to 200 MAMA pixels.  The total exposure time was
34076 sec in the clear CCD mode and 17315 sec in the far-UV mode.  Of
the 45 far-UV frames, 30 exposures were taken completely in orbital
night, for a total of 11500 sec; the remainder were taken completely
in orbital day.  Half of the night exposures spanned a zenith angle
from 61--85$\rm ^o$, and half spanned a zenith angle
from 33--48$\rm ^o$.  The day exposures spanned a zenith
angle from 47--69$\rm ^o$.  The diffuse signal in the night
exposures should have no contribution from airglow (see \S
\ref{secair}), but the day exposures clearly detected an airglow signal,
although it was noticeably smaller than the day
exposures from other programs that observed closer to the limb.

\subsection{STIS HDFS Primary Observations (HDFS-Pri)} \label{sechdfs}

In 1998 October, STIS observed a field centered on a $z=2.2$ quasar,
using many imaging and spectroscopic modes, as part of the HDFS
campaign (Williams et al.\ 2000\markcite{W00}).  The imaging modes
were the same as those used in the HDFN-Par observations
and are described in Gardner et al.\ (2000a\markcite{G00a}).  The
total exposure time was 155590 sec in the clear CCD mode, and 52124
sec in the far-UV mode.  The far-UV and clear CCD exposures each covered
a roughly square region of sky, with respective areas of 878 square
arcsec and 3009 square arcsec, both centered at
RA(J2000)=$22^h33^m38^s$ and Dec(J2000)=$\rm -60^o33\arcmin
29\arcsec$.  Dithers between far-UV frames ranged up to 135 MAMA
pixels.  Unfortunately, none of the 25 far-UV exposures
were taken completely during orbital night, and all observed far from
the zenith (the zenith angle in these exposures ranges from
74--98$\rm ^o$).  As we discuss in \S \ref{secanal}, all of these exposures
were contaminated by faint airglow that is inseparable from the FUVBG. 

\section{DATA REDUCTION} \label{secred}

Normally, multiple exposures of a field are registered and combined to
produce a summed image.  As we discuss below, we registered and summed
our far-UV images to refine our knowledge of the offsets between the far-UV
images, to verify the positional translation between the far-UV images
and the optical images which would be used for object masking, and to
allow a measurement of the far-UV flux originating in optically-detected
objects.  However, only a small portion of the detector is suitable
for a measurement of the FUVBG; to measure the FUVBG, one must mask,
for every exposure, the hot pixels, the pixels within the dark glow
region (see Figure \figschem), and optically-detected objects.
Although the dark glow can be subtracted (e.g., Gardner et al.\
2000a\markcite{G00a}; Brown et al.\ 2000\markcite{B00}), the residual
noise in glow-subtracted regions is much higher than that in regions
free of glow, and so a measurement of the FUVBG should avoid the
detector region subject to glow.  For these reasons, we found it
simpler to measure the FUVBG in individual exposures, instead of a
summed image; by examining the individual exposures, we were also able
to check for sources of systematic errors, by looking for correlations
between the measurements and observing conditions during a given
exposure (e.g., day/night, limb angle, etc.).

\subsection{Creating Object Masks} \label{secmask}

Before analyzing the far-UV images, we obtained the optical-band images
that would be used to determine object masks.  The version 1 HDFS
images and the version 2 HDFN images are available from the STScI web
site.  The STIS CCD exposures of the HDFN-Par field were reduced by
the standard reduction pipeline (including rejection of cosmic rays),
registered and summed.  On-chip $2\times2$ pixel binning produced a
plate scale of $0.1\arcsec$ pix$^{-1}$.  The plate scale for the STIS
CCD image of the HDFS-Pri field is $0.025\arcsec$ pix$^{-1}$ (via
resampling of the dithered images), and for the WFPC2 images of the
HDFN-Fol field it is $0.04\arcsec$ pix$^{-1}$.  The variation in plate
scale was unimportant for our purposes here, because we were merely
using the optical images for object masking.  Using the SExtractor
package (Bertin \& Arnouts 1996\markcite{BA96}), an object catalog was
made for each optical-band image.  For the optical images of the
HDFS-Pri and HDFN-Par fields, SExtractor was run on the summed clear
CCD STIS images.  For the optical image of the HDFN-Fol field, we made
an object catalog from the combined WFPC2 F606W and F814W images.  In
each case, the 3$\sigma$ limiting AB magnitude is $\gtrsim 29$.

We next performed a standard reduction and summation of the far-UV
data.  This process was already complete for the HDFS-Pri UV data (see
Gardner et al.\ 2000a\markcite{G00a}); we employed nearly identical
methods for the HDFN-Fol and HDFN-Par UV data.  In brief, the images
were processed via the standard pipeline, excluding the dark
subtraction, low-frequency flat field correction, and geometric
correction.  Dark frames from a six month period contemporaneous with
each science frame were processed in the same manner, and those with a
strong glow ($> 2\times 10^{-5}$ cts sec$^{-1}$ pix$^{-1}$) were
summed and fit with a cubic spline to produce a glow profile
appropriate for each frame (the shape of the glow changes slowly with
time, and so contemporaneous darks are required for this fit).  A flat
component and glow component to the dark background were then
subtracted from each science frame, and then the flat field correction
was applied.  The frames were summed with the DRIZZLE package
(Fruchter \& Hook 1998\markcite{FH98}), with registration, scaling,
and geometric correction applied to give agreement with the optical
image of each field.  This summation was iterated several times to
refine the relative offsets between individual far-UV frames.  The pixels in each
frame were weighted by EXPTIME$^2$/(SCALE$^4$ VARIANCE), where EXPTIME
is the exposure time, SCALE is the ratio of the new pixel scale to the
old pixel scale, and VARIANCE is the dark count variance.  The weight
map also included a hot pixel mask.  Because the 
dark count variance is the product of the dark rate and the exposure
time in a given pixel, our weight scales linearly with the ratio of the
exposure time to the dark rate; equivalently, the exposures were weighted
by the square of the signal-to-noise (S/N) ratio for sources
that were fainter than the background, thus optimizing the summation
to account for the temporal and spatial variations in the dark
background.  The statistical errors in each final drizzled far-UV
image, for objects below the background, were given by the square root
of the final drizzled weights map.  In a separate paper (Gardner et
al.\ 2000b\markcite{G00b}), these registered images will also be used for
an analysis of the far-UV number counts.

\bigskip

\hskip -0.2in
\parbox{3.0in}{\epsfxsize=3.5in \epsfbox{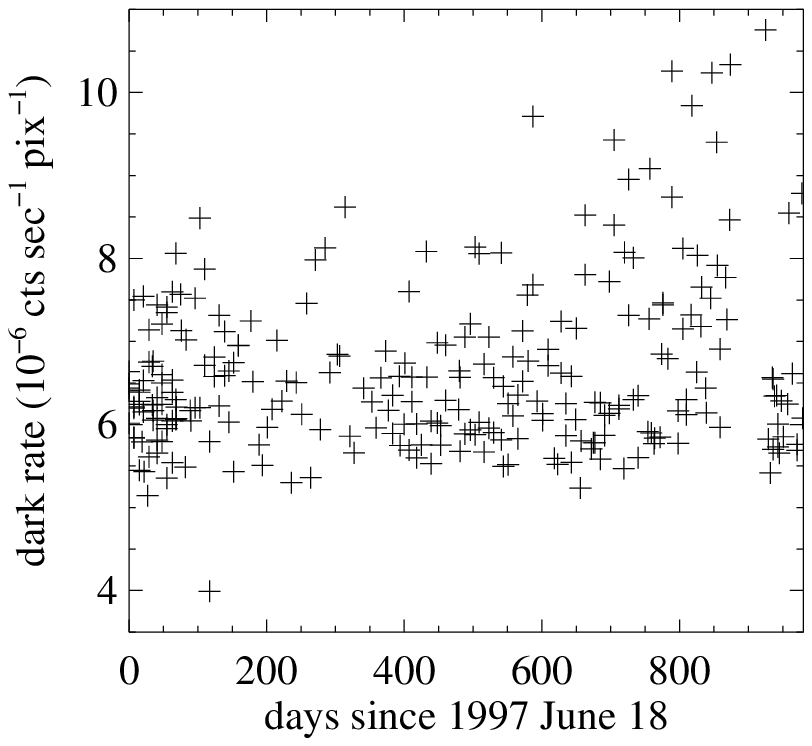}} \vskip 0.1in

\centerline{\parbox{3.5in}{\small {\sc Fig.~\figdark--} These 268 dark
exposures (1380 sec each) demonstrate the stability of the STIS dark
rate in the region free of glow, hot pixels, and occulted pixels (see
Figure \figschem).  The mean dark rate for these exposures 6.6$\times
10^{-6}$ cts sec$^{-1}$ pix$^{-1}$ with an rms of $1.0 \times 10^{-6}$
cts sec$^{-1}$ pix$^{-1}$.  The late gap in the history is due to the
safing of HST and subsequent servicing mission.}} \addtocounter{figure}{1}

\bigskip

Once all of the far-UV frames were drizzled into a summed image for
each of the three datasets, the BLOT package made it trivial to determine
the object mask for each individual far-UV frame.  Using BLOT
(available with DRIZZLE as part of the IRAF DITHER package), the
SExtractor segmentation map for each optical image was
reverse-drizzled back to each individual far-UV frame, thus creating a
tailored set of object masks for all of our far-UV frames.  When
combined with the detector mask shown in Figure \figschem, the
remaining set of unmasked pixels in each exposure were suitable for
measurements of the FUVBG.  However, to measure the FUVBG, we first
determined the dark rate in each frame.

\subsection{Dark Rate Determination} \label{secdark}

As discussed in \S \ref{secobs}, the STIS far-UV camera has a very low
but spatially and temporally variable dark rate that correlates with
the detector temperature.  At the time of this writing, there were 268
normal dark frames available in the STScI archive, 187 of which were
taken during the time frame spanned by our science observations.  The
dark variation appears as a glow in the upper left hand quadrant of
the detector (see Figure \figschem); when this glow is strong, it can
peak at 20$\times$ the low dark rate that exists outside of the glow.
Pixels on the detector that are far from the glow have a much
more stable dark rate (see Figure \figdark), of 6.6$\times 10^{-6}$
cts sec$^{-1}$ pix$^{-1}$, with an rms of $1.0 \times 10^{-6}$ cts
sec$^{-1}$ pix$^{-1}$ between dark frames.

To determine the dark rate in an individual frame, we took advantage
of the fact that the aperture mask for the far-UV camera occults the
bottom $\sim 15$ pixels in STIS images.  By comparing the dark rate in
rows 2--11 with the dark rate in that part of the detector appropriate
for measuring the FUVBG (see Figure \figschem), we found that the
occulted region consistently had a dark rate 1.1$\times$ higher, due to edge
effects with the microchannel plate signal processing.  In our science
frames, we assumed that the appropriately scaled rate in the occulted
region gave the dark rate where we measured the FUVBG.

\section{THE DIFFUSE FAR-UV BACKGROUND} \label{secanal}

For each far-UV exposure in all three datasets, we used IDL to
calculate the net FUVBG signal in the unmasked portion of the
detector.  The mask included astronomical objects, hot pixels,
occulted pixels, the dark glow region, and any pixel within 20 pixels
of the image edge; the dark rate was determined from the occulted
portion of the detector.  We used the science frames that were reduced
via the standard pipeline, excluding the dark subtraction,
low-frequency flat field correction, and geometric correction; i.e.,
we did not include the later corrections explained in \S
\ref{secmask}.  Surprisingly, the FUVBG signal measured in each
science frame varied much more than the expected variation from
Poisson statistics (which included both the statistical uncertainty in
the signal and the dark rate determination).  The signal did not
correlate with the strength of the dark glow, thus ruling out an
incomplete masking of the glow region.  Instead, the signal correlated
with the fraction of the observation taken during orbital day, as
shown in Figure \figgeo.

The HDFS-Pri and HDFN-Fol UV data were all taken near the limb (as
opposed to HDFN-Par UV data - see \S \ref{secpar}).  These two
near-limb datasets, taken together, show a strong correlation between
our attempted FUVBG measurement and that fraction of the exposure
spent in orbital day: the Pearson's linear correlation coefficient is
0.77.  Fitting a line to these points and errors gave a y-intercept of
$379 \pm 57$ ph cm$^{-2}$ sec$^{-1}$ ster$^{-1}$ \AA$^{-1}$ for
near-limb observations at complete orbital night.  However, the
variation in the airglow was clearly affected by more
parameters than the ``day fraction'' of the exposure, because the
points taken in complete orbital day show a large degree of variation:
the rms of the points in complete orbital day is 576 ph cm$^{-2}$
sec$^{-1}$ ster$^{-1}$ \AA$^{-1}$, while the statistical errors are
half that size.  Much of this variation is likely due to the
variations in limb angle.  Thus, the linear fit shown in Figure
\figgeo\ is simply a rough approximation of the trend in these data.
The HDFN-Par UV observations are not included in Figure \figgeo,
because they were all taken in either complete orbital
day or night with very short exposure times, and thus large statistical
uncertainties.

Two of the programs discussed herein include data taken in complete
orbital night: the HDFN-Fol and the HDFN-Par programs.  For those
night exposures, we calculated a weighted sum of the diffuse
dark-subtracted signal in each frame, where the weight is the product
of the number of exposed pixels and the exposure time, thus maximizing
the S/N and minimizing the contribution from brief exposures with few
unmasked pixels.  The measurement from the night HDFN-Par frames is
$296 \pm 135$ ph cm$^{-2}$ sec$^{-1}$ ster$^{-1}$ \AA$^{-1}$, and the
measurement from the night HDFN-Fol frames is $712 \pm 157$ ph
cm$^{-2}$ sec$^{-1}$ ster$^{-1}$ \AA$^{-1}$; together, the entire set
of night frames yields a measurement of $501 \pm 103$ ph cm$^{-2}$
sec$^{-1}$ ster$^{-1}$ \AA$^{-1}$.  As shown in Table~\tabobs, both of
these fields sample Galactic sight-lines with very low $N_{HI}$ and
extinction (the excluded dayglow-contaminated HDFS-Pri data samples a
sight-line with somewhat higher hydrogen column and extinction).  Our
night measurement of the FUVBG is in good agreement with the earlier
measurements of 300 ph cm$^{-2}$ sec$^{-1}$ ster$^{-1}$ \AA$^{-1}$,
calculated by extrapolating to $N_{HI}=0$ (Bowyer 1991\markcite{B91}
and references therein).  Although we could not determine from these
data alone if there was a contribution from airglow in our
night measurement, many experiments have shown that the night
airglow should be several orders of magnitude below the
STIS dark rate.  We discuss possible contamination from airglow
in the next section.

\section{AIRGLOW} \label{secair}

The STIS crystal quartz filter blocks emission shortward of 1450~\AA\
very effectively, and the long wavelength cutoff of the far-UV
bandpass is due to detector sensitivity; the varying signal detected
in our near-limb day observations probably originated in airglow
within the 1450--1900~\AA\ bandpass.  The strongest
airglow line in this wavelength range is \ion{N}{1} $\lambda
1493$~\AA\ (Eastes et al.\ 1985\markcite{E85}; Budzien, Feldman, \&
Conway 1994\markcite{BFC94}).  Viewing at an altitude of $\sim 200$ km
and $90 \rm ^o$ to the zenith (in 1980, near the solar maximum) Eastes
et al.\ (1985\markcite{E85}) observed $\sim 1000$ R of \ion{N}{1}
dayglow emission.  For atomic nitrogen at an altitude of
200--600 km, the scale height is $k T / (M g) \approx 60 $ km.  STIS
observes at an altitude of 600 km, and the observations described
herein were also all taken within a few years of solar maximum; at 6.7
scale heights above the measurements of Eastes et
al. (1985\markcite{E85}), STIS might expect to observe about one
rayleigh of \ion{N}{1}, which translates to $6 \times 10^{-7}$ cts
sec$^{-1}$ pix$^{-1}$.  Instead, from Figure \figgeo, we see that
exposures in complete orbital day produced a net dark-subtracted
signal that was 2--8 times higher.  Thus, the STIS signal may reflect
the combined contribution from several weaker sky emission lines,
and/or the fact that the optically-thin \ion{N}{1} emission is not
accurately described by the approximation we have used here.

Eastes et al.\ (1985\markcite{E85}) showed that there are several N$_2$
Lyman-Birge-Hopfield (LBH) emission lines in the 1450--1900~\AA\
bandpass.  These lines are individually somewhat weaker than the
\ion{N}{1} emission, but their summed signal is several thousand
rayleigh at an altitude of 200~km.  However, the scale height of N$_2$
is only 30~km, so the contribution from the LBH emission should be
negligible at 600~km.

Observing near solar maximum from the Space Shuttle at an altitude of
358 km, Waller et al.\ (1995\markcite{W95}) also found significant
dayglow emission in a bandpass similar to the STIS far-UV bandpass;
they attributed their 150 R of near-limb emission to both \ion{N}{1}
and N$_2$.  At 2.6 \ion{N}{1} scale heights and 5.3 N$_2$ scale
heights above the Eastes et al.\ (1985\markcite{E85}) measurements,
one would expect 75 R of \ion{N}{1} emission and $\sim$10 R of N$_2$
emission.  Thus, Waller et al.\ (1995\markcite{W95}) observed about
twice as much near-limb dayglow as one might expect from extrapolating
the measurements of Eastes et al.\ (1985\markcite{E85}), and the far-UV
dayglow measurements at 200 km, 358 km, and 600 km all agree within an
order of magnitude.

\bigskip
 
\hskip -0.2in
\parbox{3.0in}{\epsfxsize=3.5in \epsfbox{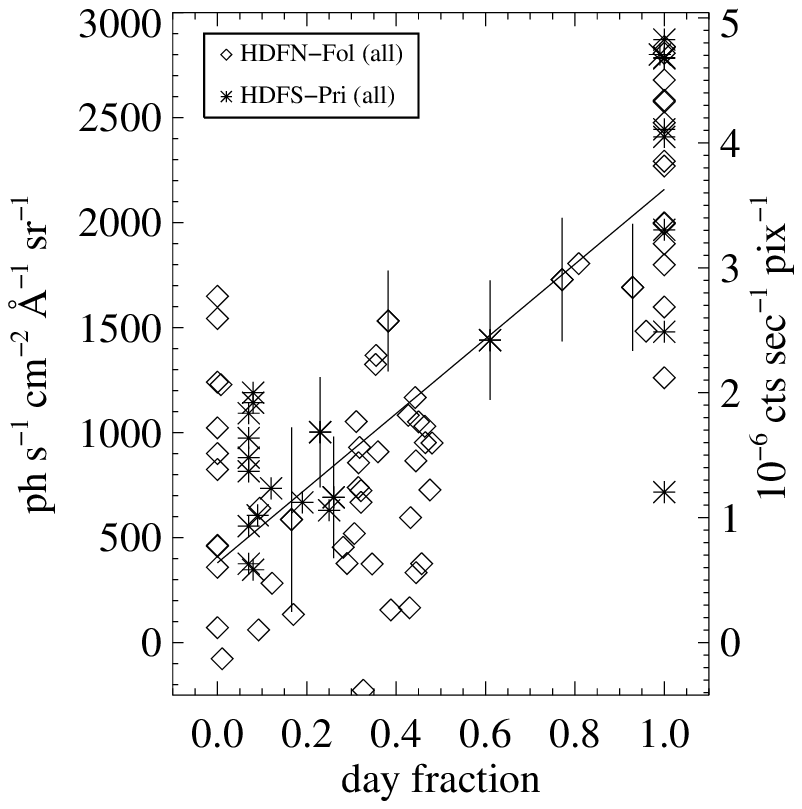}} \vskip 0.1in

\centerline{\parbox{3.5in}{\small {\sc Fig.~\figgeo--} Our
measurements of the FUVBG from exposures near the limb,
versus that fraction of the exposure
taken during orbital day.  Measurements from the HDFN-Fol UV data
({\it diamonds}) and the HDFS-Pri UV data ({\it asterisks}) are shown
with a representative sample of their statistical uncertainties.  The
variation among these measurements is larger than that expected from the
statistics, and correlates with the
fraction of the exposure taken in orbital day.  The line shows the
uncertainty-weighted fit to these data, with a
y-intercept of $373 \pm 67$.  For data
taken in complete orbital night, there should be no signal from
airglow (see \S \ref{secair}).  Note
that the net diffuse signal that we measured was always less than
the STIS dark rate (compare the right-hand y-axis with Figure
\figdark), but it was a nonnegligible contribution to the total
background in near-limb day observations. }} \addtocounter{figure}{1}

\bigskip

Previous observations of far-UV airglow during orbital night showed
signals that were far below those observed during the day.  For
example, observing near the limb at an altitude of 330 km, Morrison et
al.\ (1992\markcite{R92}) easily detected a few tenths of a rayleigh
in \ion{O}{1} emission lines at 1301 and 1356~\AA, but did not detect
the much weaker \ion{N}{1} 1493~\AA\ emission, nor any other line that
would produce significant airglow in the STIS bandpass.  Morrison et
al.\ (1992\markcite{R92}) observed near solar minimum, while STIS
observed near solar maximum, but the STIS night observations were
taken at an altitude that is 4 scale heights above the measurements of
Morrison et al.\ (1992\markcite{R92}); the STIS observations should be
completely insensitive to \ion{N}{1} airglow, for any
zenith angle $< 95^{\rm o}$ (at an HST altitude of 600 km, a sight
line along a zenith angle of 95$^{\rm o}$ only intersects atmospheric
heights > 575 km).  However, it is troubling that the HDFN-Fol
measurement of the FUVBG (taken over a zenith angle of 82--95$\rm ^o$)
is considerably higher than the HDFN-Par measurement (taken at in two
zenith angle ranges of 61--85$\rm^o$ and 33--48$\rm ^o$).  We show the
individual night measurements of the FUVBG versus limb angle in
Figure~\figlimb.  The statistics are poor, but there is a slight trend
for an increasing signal as one moves to higher limb angle; the
Pearson's linear correlation coefficient for the signal versus zenith
angle, for the points above a zenith angle of 60$\rm ^o$, is 0.33 (it
is only 0.13 if the points at all zenith angles are included, but the
nonlinear increase in signal should mainly occur as one approaches the
limb).  We stress that experiments at lower altitudes indicate that
our night FUVBG measurements should have no significant contribution
from airglow.  A well-designed STIS measurement of the
FUVBG should be taken in TIMETAG mode, which would allow the
extraction of night-only data from long exposures, and demonstrate the
sensitivity (or lack thereof) to zenith angle.

\section{EMISSION FROM RESOLVED OBJECTS} \label{secgal}

Because our object masks were derived from very deep optical images,
we masked all objects to an optical AB magnitude of 29 in our
far-UV images, giving a measure of the truly diffuse far-UV emission.
In a subsequent paper, we will discuss in detail the far-UV emission
from these objects, but it is worth noting here their summed
contribution to these fields.  For this measurement, we used the
drizzled sum of far-UV exposures in each field (see \S \ref{secmask}),
including those regions affected by the dark glow.  A glow profile was
subtracted from the individual frames before drizzling, and any
residual dark glow was subtracted as part of the local sky value
for each object.  The increased noise in the regions of higher dark
rate does not significantly reduce the S/N ratio of this measurement.
Instead, the dominant source of error is the incomplete sampling of
the number counts, due to the small sky coverage.

To determine the flux for each object, we summed the countrate within
the area for each object, using the SExtractor segmentation map to
define that area, and then subtracted the mean local background within
a $6\arcsec \times 6 \arcsec$ box. Our method was the same as that
used for the HDFS catalog (Gardner et al.\ 2000a\markcite{G00a}), which
includes a detailed description of the procedure.  In the HDFN-Par UV
data, the emission originating in optically-detected objects is $94
\pm 14$ ph cm$^{-2}$ sec$^{-1}$ ster$^{-1}$ \AA$^{-1}$.  In the
HDFN-Fol UV data, we measured $60 \pm 4$ ph cm$^{-2}$ sec$^{-1}$
ster$^{-1}$ \AA$^{-1}$ in optically-detected objects, although the
measurement drops to $45 \pm 4$ ph cm$^{-2}$ sec$^{-1}$ ster$^{-1}$
\AA$^{-1}$ if one excludes a blue Galactic star from the field.  
In the HDFS-Pri data, we measured $26 \pm 5$ ph cm$^{-2}$ sec$^{-1}$
ster$^{-1}$ \AA$^{-1}$ in optically-detected objects, excluding the
very bright target $z=2.2$ quasar in the center of the field.  The
uncertainty in these measurements was dominated by the inadequate
sampling of the UV number counts, which exceeded the errors 
quoted here for photon counting statistics.  The contribution from faint
objects in these fields shows that galaxies will make a nonnegligible
(though not dominant)
contribution to a measurement of the FUVBG if they are not properly
masked or subtracted via an assumed UV number counts distribution.

\section{DISCUSSION} \label{secdis}

Restricting our data to that obtained in orbital night, we measured a
diffuse far-UV background of $501 \pm 103$ ph cm$^{-2}$ sec$^{-1}$
ster$^{-1}$ \AA$^{-1}$.  Our $5 \sigma$ detection of the FUVBG, a
signal that is $\sim$10\% of the STIS dark rate, is in reasonable
agreement with earlier measurements at low $N_{HI}$.  However, the
unique aspect to STIS measurements of the FUVBG is that STIS can
measure a truly diffuse signal, because high-resolution coincident
imaging in the far-UV and optical bandpasses allows masking of all
objects down to a very faint magnitude.

The night data used for our measurement of the FUVBG were taken in two
fields at high Galactic latitude, separated by about 8.5 arcmin.  The
measurement in the HDFN-Par field was $296 \pm 135$ ph cm$^{-2}$
sec$^{-1}$ ster$^{-1}$ \AA$^{-1}$, and the measurement in the HDFN-Fol
field was $712 \pm 157$ ph cm$^{-2}$ sec$^{-1}$ ster$^{-1}$
\AA$^{-1}$.  The HDFN-Fol field was observed much closer to the
Earth's limb than the HDFN-Par field, but the night exposures should
show no measurable contribution from airglow.  Barring
systematic errors, the difference between the two fields may reflect a
true patchiness in the FUVBG, whether it originates in scattering from
Galactic dust or an extragalactic source.  Note that the Galactic
$E(B-V)$ along each line of sight is very low (0.01~mag).

\bigskip

\hskip -0.2in
\parbox{3.0in}{\epsfxsize=3.5in \epsfbox{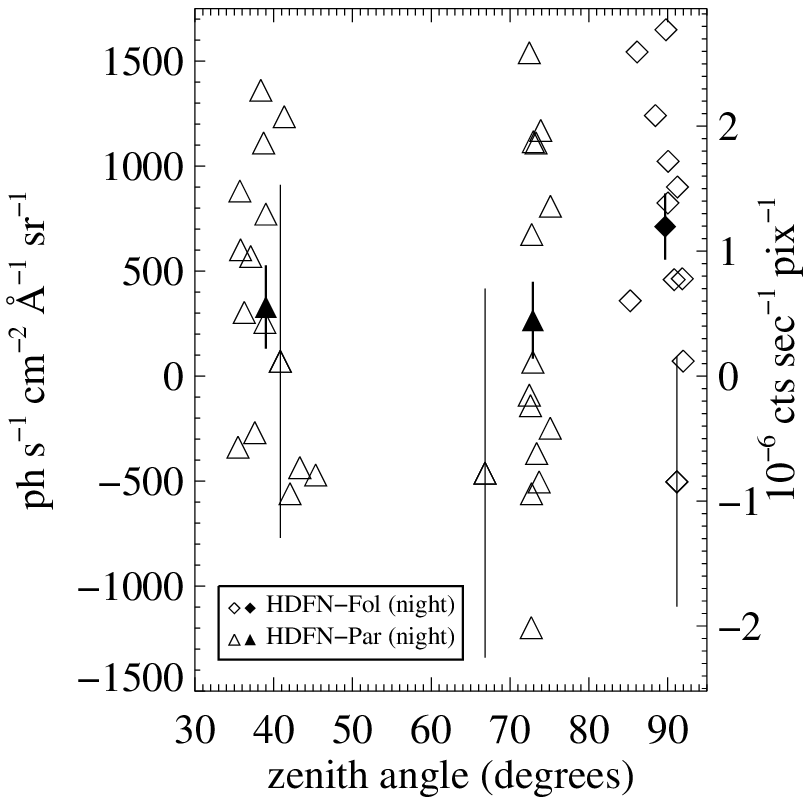}} \vskip 0.1in

\centerline{\parbox{3.5in}{\small {\sc Fig.~\figlimb--} Our
measurements of the FUVBG, as a function of the average zenith angle
during the exposure, using only those exposures taken in complete
orbital night.  Individual measurements from the HDFN-Fol data
({\it open diamonds}) and the HDFN-Par data ({\it open triangles})
are shown with a representative sample of their associated
uncertainties.  The filled symbols show the sum of the measurements
near 39$\rm ^o$, 73$\rm ^o$, and 90$\rm ^o$, with their associated
uncertainties; there is a weak trend for increasing signal at high
zenith angle (near the limb, which is at 114$\rm ^o$), but the known
night airglow in this bandpass should be completely
negligible at the HST altitude of 600 km.  The sum of the data taken at
a zenith angle less than 80$\rm ^o$ is $296 \pm 135$ ph cm$^{-2}$
sec$^{-1}$ ster$^{-1}$ \AA$^{-1}$, and the sum of the data taken at a
zenith angle greater than 80$\rm ^o$ is $712 \pm 157$ ph cm$^{-2}$
sec$^{-1}$ ster$^{-1}$ \AA$^{-1}$.  
}} \addtocounter{figure}{1}

\medskip

To measure the FUVBG with high statistical certainty, future STIS
observations should be obtained in TIMETAG mode, to exclude day data,
and to ensure that the night data shows no correlation with zenith
angle.  The target should be a reasonably empty portion of a field
with deep optical imaging, such as one of the fields described herein.
If the STIS cooling system to be installed in HST Servicing Mission 3B
is as effective as expected in eliminating dark glow, a much larger
fraction of the detector area will be available for sensitive
measurements.  Off-nominal positioning of the entrance aperture mask
could also increase the shadowed detector area available for
concurrent monitoring of the dark rate, though HST operating
constraints may not permit this.

Prior to our analysis, the crystal quartz filter on the STIS far-UV
camera was thought to produce a bandpass that was completely
insensitive to both zodiacal emission and airglow.  Because the
dark rate in this camera is so low, the low sky enables photometric
measurements of very faint extended UV sources (see, e.g., Brown et
al.\ 2000\markcite{B00}).  However, our analysis showed that while the
sky signal is always lower than the dark rate, it may be
nonnegligible.  The dark glow can increase the background signal by a
factor of 20 in sections of the detector, but the sky can reach levels
comparable to the minimum dark rate; the sky can thus increase the
background signal by a factor of two in the darkest parts of the
detector during near-limb day observations.  Potential STIS observers
should take this sky signal into account when planning their S/N
estimates, if they will be observing near the limb (e.g., in the HST
Continuous View Zone).

\acknowledgments Support for this work was provided by NASA through
the STIS GTO team funding.  TMB acknowledges support at GSFC by
NAS~5-6499D.  We are grateful to P. D. Feldman for discussions of UV
airglow.  We are grateful to R. Thompson for allowing
parallel STIS observations to the NICMOS HDF imaging program.

\end{document}